\documentclass[aps,pra,reprint,twocolumn,superscriptaddress]{revtex4-1}

\usepackage{amssymb}
\usepackage{graphicx}
\usepackage{amsmath}
\usepackage{bbm}
\newcommand{\Tr}{{\mathrm{Tr}}}
\usepackage{nonfloat}
\usepackage[pdftex]{hyperref}

\begin{document}

\title{Classical capacity of Gaussian thermal memory channels}
\author{G. De Palma}
\affiliation{NEST, Scuola Normale Superiore and Istituto Nanoscienze-CNR, I-56127 Pisa,
Italy.}
\affiliation{INFN, Pisa, Italy}
\author{A. Mari}
\affiliation{NEST, Scuola Normale Superiore and Istituto Nanoscienze-CNR, I-56127 Pisa,
Italy.}
\author{V. Giovannetti}
\affiliation{NEST, Scuola Normale Superiore and Istituto Nanoscienze-CNR, I-56127 Pisa,
Italy.}

\begin{abstract}
The classical capacity of phase-invariant Gaussian channels has been recently determined under the assumption that such channels are memoryless. In this work we generalize this result by deriving the classical capacity of a model of quantum memory channel, in which the output states depend on the previous input states. In particular we extend the analysis of [C. Lupo, {\it et al.}, PRL and PRA (2010)] from quantum limited channels to thermal attenuators and thermal amplifiers. Our result applies in many situations in which the physical communication channel is affected by nonzero memory and by thermal noise.
\end{abstract}

\maketitle

\section{Introduction}

Given a physical device acting as a quantum communication channel \cite{caves,holevo}, an important problem in quantum information theory is to
determine the optimal rate of classical information that can be sent through the channel assuming that one is allowed to use arbitrary
quantum encoding and decoding strategies possibly involving multiple uses of the transmission line ({\it channel uses}). The maximum achievable rate is the {\it classical capacity} associated to the quantum channel \cite{holevo,holevowerner,schumacher}.  A simple closed formula for this quantity
does not exist, since typically it is not easy to see whether entangled input states will improve the communication rate. Still it is possible to prove \cite{holevo} that
if no memory effects are tampering the communication line (i.e. if the noise affecting the communication acts identically and independently on subsequent channel uses)
the classical capacity of the setup can be expressed as the following limit
\begin{equation}
C(\Phi)=\lim_{n\to\infty}\frac{1}{n}\chi\left(\Phi^{\otimes n}\right)\;,\label{capacityformula}
\end{equation}
where $\Phi$ is the (completely positive, trace preserving) mapping characterizing the input-output relations of a single channel use, and where
 $\chi\left(\Phi^{\otimes n}\right)$ is the Holevo information of $n$  channel uses, which is defined through the identity
\begin{equation}
\chi(\Phi)=\sup_\mu\left[ S\left(\int\Phi(\rho)d\mu(\rho)\right)-\int S\left(\Phi(\rho)\right)d\mu(\rho)\right]\;,
\end{equation}
 the supremum being taken over all probability measures $\mu$ on the space of the density matrices $\rho$ of the system, and $S$ being the von Neumann entropy, i.e.
$S(\rho)=-\mathrm{Tr}\left[\rho\ln\rho\right]$.

Most real communication media are based on electromagnetic signals and are well described within the framework of quantum Gaussian channels \cite{braunstein, gauss1,gauss2}.
The most relevant class is constituted by phase-invariant channels like attenuators and amplifiers. Such channels reduce or increase the amplitude of the signal
and, at the same time, they add a certain amount of Gaussian noise which depends on the vacuum or thermal fluctuations of the environment.
Recently the proof of the minimum output entropy conjecture \cite{conj1,conj2} has allowed the determination of the exact classical capacities of these channels \cite{CC} and the respective strong converse theorems \cite{strong}, under the crucial assumption of
their memoryless behavior. One of the key points of the proof is the additivity of the $\chi$ capacity of a memoryless phase-invariant gaussian channel:
\begin{equation}
\chi\left(\Phi^{\otimes n}\right)=n\chi(\Phi)\;,
\end{equation}
\begin{figure}[t]
\includegraphics[width=1 \columnwidth]{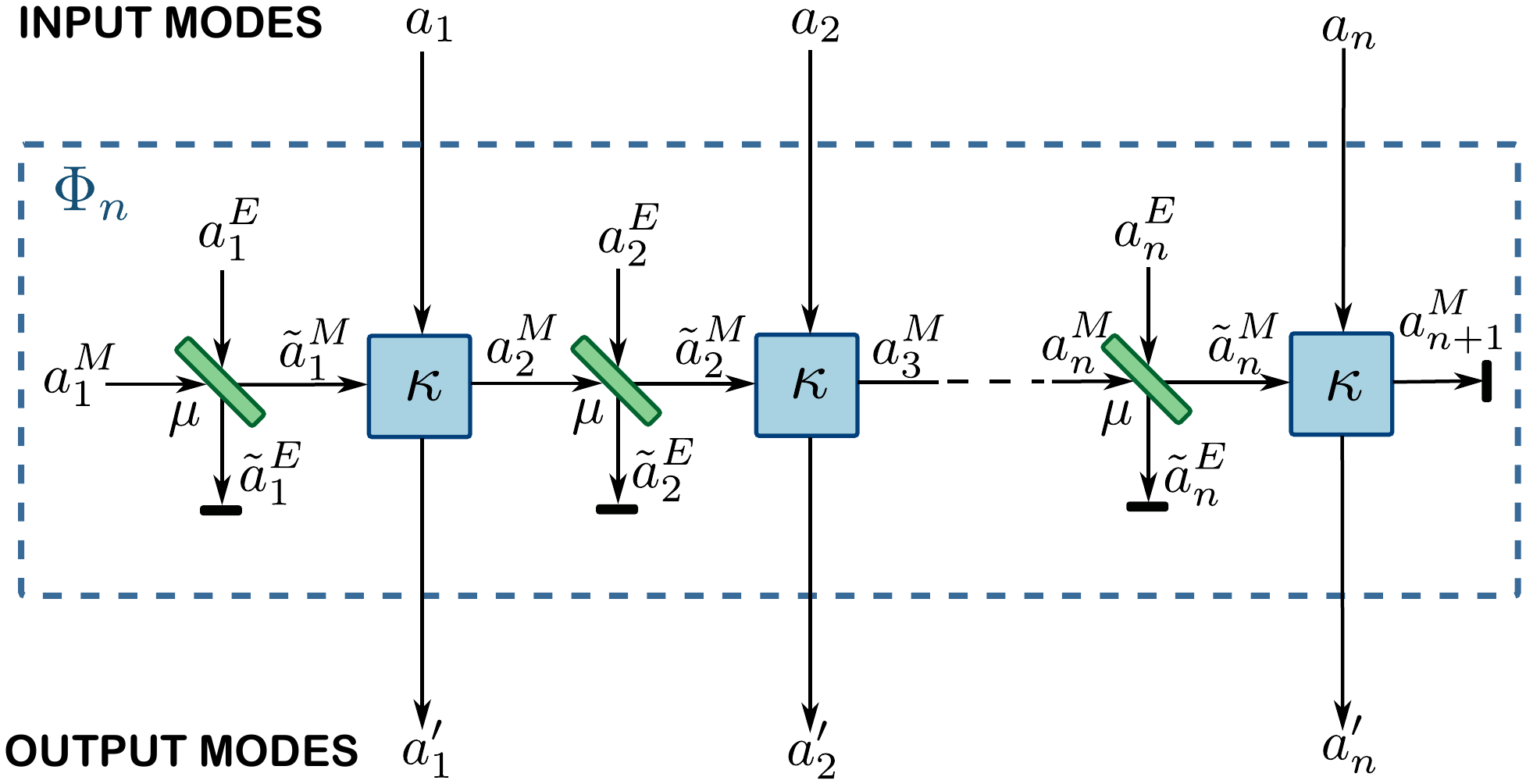}
\caption{(Color online): Schematic description of a Gaussian memory channel $\Phi_n$ which is iterated $n$ times. The application of a the memory channel to $n$ successive input modes $a_1,\  \dots,a_n $ is described by $n$ phase-insensitive channels $\mathcal E_\kappa$ (thermal attenuators or amplifiers) where each of them is coupled to a Gaussian thermal environment and to a memory mode. The initial memory mode $a^{M}_1$ travels horizontally and correlates the output signals with the previous input signals. A  beam-splitter of transmissivity $\mu$ is used to tune the memory effect of the channel. For $\mu=1$ the memory mode is perfectly preserved while for $\mu=0$ the channel becomes memoryless. A reasonable choice for the initial state of the memory mode is a Gaussian thermal state in equilibrium with the environment, {\it i.e.}\ we make the identification $a^{M}_1=a^{E}_0$. The final state of the memory is assumed to be inaccessible and is traced out. }\label{model}
\end{figure}
which trivializes the limit in \eqref{capacityformula}.
Realistic communication lines however, if used at high rates (larger than the relaxation time of the environment), may exhibit
memory effects in which the output states are influenced by the previous input signals \cite{memoryrev,memory0, exp1, exp2}.  In other words, the noise introduced by the channel instead of being independent and identically distributed can  be correlated
with the previous input states preventing one from expressing the input-output mapping  of $n$ successive channel uses as a simple tensor product $\Phi^{\otimes n}$
 and hence from using
 Eq.~(\ref{capacityformula}). As a matter of fact since the capacity is defined asymptotically in the limit of many repeated channel uses,  memory effects will affect the optimal information rate and the optimal coding strategies.
A characterization of {\it quantum memory channels} can be found in Ref.s\ \cite{memory1,memory2,memory3}, while generalizations to infinite dimensional bosonic systems are considered in  \cite{memory4,CERF2,lupo1,lupo2,OLEG,CERFA,CERF1}.

Here we elaborate on the model of (zero temperature) attenuators and amplifiers with memory effects that was introduced in Ref.s \ \cite{lupo1, lupo2} where, in the case of a quantum limited attenuator, the capacity was explicitly determined. In this work we generalize this model to thermal attenuators and thermal amplifiers and we derive the corresponding classical capacities, extending the previous results obtained in the memoryless scenario \cite{CC}. We have also considered the case of the additive noise channel, viewed as a particular limit of an attenuator with large transmissivity and large thermal noise. This limit is essentially equivalent to the model
considered in \cite{additiven1,CERFA}, and we have shown that the only effect of the memory is a redistribution of the added noise.
An interesting fact which emerges from our analysis is the presence of a critical environmental temperature which strongly affects
 the distribution of the input
energy  among the various modes of the model. In particular  for temperatures larger than the critical one, only the modes which have a sufficiently high effective  transmissivity are allowed to
contribute to the signaling process, the remaining one being forced to carry no energy nor information.

Given a quantum channel the associated unitary dilation is not unique and one can imagine different models for memory effects.
Nonetheless our paradigm is expected to cover many real devices like optical fibers  \cite{fibers,waveguides},  microwave systems \cite{microwaves}, $THz$ lasers \cite{tredicucci}, free space communication \cite{freespace}, {\it etc}..
All physical implementations are known to exhibit time delay and memory effects whenever used at sufficiently high repetition rates. Moreover, especially in microwave and electrical channels, thermal noise is not
negligible and will affect the classical capacity. In general, our analysis applies to any physical realization of quantum channels in which memory effects and thermal noise are simultaneously present.

We begin in Sec.~\ref{sec:gaus} by recalling some basic facts about the memory channel model of Ref.s \cite{gauss1,gauss2}. In particular we describe its
normal mode decomposition which allows one to express the associated mapping as a tensor product of  not necessarily identical single mode transformations.
In Sec.~\ref{sec:capac} instead we compute the classical capacity of the setup and discuss some special cases, while in Sec.~\ref{sec:lagr} we analyze how the distribution of the input energy among the various modes is affected by the  presence of a thermal environment.
Conclusions and perspectives are provided in Sec.~\ref{sec:conc}.

\section{Gaussian memory channels}\label{sec:gaus}

In this section we review the model of {\it Gaussian memory channels} introduced in Ref.s\ \cite{lupo1,lupo2}. We closely  follow their analysis showing that this memory channel can be reduced to a collection of memoryless channels by some appropriate
encoding and decoding unitary operations.

\subsection{ Quantum attenuators and amplifiers}
The  building blocks of our analysis are single mode quantum attenuators and amplifiers \cite{gauss1,gauss2}. Let us consider a continuous variable bosonic system \cite{braunstein} described by the creation and annihilation
operators $a$ and $a^\dag$ and another mode described by $a^E$ and $a^{E \dag}$ associated to the environment. We focus on two important Gaussian unitaries,
\begin{subequations}
\begin{eqnarray}
U_\kappa&=&e^{\arctan\sqrt{\frac{1-\kappa}{\kappa}}\left(a\, a^{E\dag} - a^\dag a^E\right)}\;,\quad {\rm for}\;  \kappa  \in [0,1],\\
U_\kappa &=&e^{\mathrm{arctanh}\sqrt{\frac{\kappa-1}{\kappa}}\left(a^\dag a^{E \dag} - a\, a^E \right)}\;,\quad {\rm for}\;   \kappa >1,
\end{eqnarray}
\end{subequations}
corresponding to the beam-splitter and two-mode squeezing operations, respectively. Their action on the annihilation operator is
\begin{subequations}
\begin{eqnarray}
U_\kappa ^\dag a U_\kappa &=&\sqrt{\kappa}\; a - \sqrt{1- \kappa}\; a^E\;,\quad {\rm for}\;  \kappa  \in [0,1], \label{U1} \\
U_\kappa^\dag a U_\kappa&=&\sqrt{\kappa}\; a + \sqrt{\kappa-1}\; a^{E \dag}\;,\quad {\rm for}\;   \kappa >1. \label{U2}
\end{eqnarray}
\end{subequations}
If the environment is in a Gaussian thermal state $\rho_E= e^{- \beta \hbar \omega a^{E \dag} a^E}\left/ \mathrm{Tr}\left[ e^{- \beta \omega \hbar a^{E \dag} a^E}\right]\right.$ with mean photon number $N=\mathrm{Tr}\left[a^{E \dag} a^E \rho\right]= \left(e^{\beta \hbar \omega}-1\right)^{-1}$, applying the unitaries \eqref{U1} and \eqref{U2} and tracing out the environment, we get

\begin{equation}
\mathcal E_\kappa(\rho)= \mathrm{Tr}_E \left[ U_\kappa (\rho  \otimes \rho_E) U_\kappa^\dag\right].
\end{equation}
This generates two different phase-insensitive channels depending on whether $\kappa$ is less or larger than 1. For $\kappa \in [0,1]$  the channel corresponds to a thermal attenuator, while for
$\kappa >1$ the channel is a thermal amplifier. In both cases the classical capacity has been recently determined in \cite{CC}. Under the input energy constraint  $\mathrm{Tr}\left[a^\dag a \rho\right] \leqslant E $, the capacities of the   attenuator and of the amplifier are obtainable via a Gaussian encoding and are given by \cite{CC} (in nats for channel use):
\begin{subequations}
\begin{align}
C_{\kappa\in [0,1]}&=g[\kappa E + (1-\kappa) N] - g[(1-\kappa) N], \label{capbs}\\
C_{\kappa>1}&=g[\kappa E + (\kappa-1) (N+1)]  - g[(\kappa-1) (N+1)]\label{capamp}\;,
\end{align}
\end{subequations}
where $g(x)=(x+1)\ln(x+1)-x\ln(x)$.

\subsection{Gaussian memory channels}

In order to include memory effects we follow the model introduced in \cite{lupo1, lupo2} and schematically shown in Fig.\ \ref{model}.
In addition to the degrees of freedom of the system and of the thermal environment we introduce a ``memory'' described by the bosonic operators $a^M$ and $a^{M \dag}$.
The channel acts in the following way: as a first step the memory is mixed with the environment via a beam-splitter of transmissivity $\mu$,
\begin{eqnarray}
\tilde a^M&=&\sqrt{\mu}\; a^M + \sqrt{1- \mu}\; a^E.
\end{eqnarray}
The outcome state is used as an effective environment for the quantum attenuator or alternatively the quantum amplifier. More precisely, the second step consists in applying the unitary \eqref{U1} or \eqref{U2} to the product state of the system and of the effective environment,
 \begin{subequations}
 \begin{eqnarray}
 a' &=&\sqrt{\kappa}\; a - \sqrt{1- \kappa}\; \tilde a^M, \quad  \kappa  \in [0,1], \\
 a' &=&\sqrt{\kappa}\; a + \sqrt{\kappa-1}\; \tilde a^{M \dag}, \quad  \kappa>1.
\end{eqnarray}
\end{subequations}
The second port of the attenuator or amplifier is given by the corresponding complementary channel,
 \begin{subequations}
 \begin{eqnarray}
 a^{M'} &=&\sqrt{\kappa} \;\tilde a^M + \sqrt{1- \kappa}\; a, \quad  \kappa  \in [0,1], \\
 a^{M'} &=&\sqrt{\kappa} \;\tilde a^M + \sqrt{\kappa-1}\; \tilde a^\dag, \quad  \kappa>1.
\end{eqnarray}
\end{subequations}
The complementary mode described by the annihilation operator $a^{M'}$ contains a fraction of the amplitudes of the input state, and represents the updated state of the memory, {\it i.e.} in the next use of the channel,
the mode $a^{M'}$ will play the role of the previous memory operator $a^M$. Once the initial states of the memory and
of the environment are specified, the action of the channel after $j$ uses is completely determined and can be computed
recursively. The explicit formula for the $j$th output mode can be found in \cite{lupo2} and is not repeated here.
What is important is just the structure of the equations
\begin{subequations}
\begin{align}
a_j'= \sum_{h=1}^{j-1} A_{jh} \, a_h - \sum_{h=0}^{j} E_{jh} \, a^E_h \, , \quad & \kappa  \in [0,1] \, , \label{outputmodes}\\
a_j' = \sum_{h=1}^{j-1}  A_{jh} \, a_h +\sum_{h=0}^{j}  E_{jh} \, a_h^{E \dag} \, ,
\quad & \kappa > 1 \, ,
\end{align}
\end{subequations}
where $A$, $E$, are real matrices and the initial state of the memory has been identified with an additional mode of the environment $a^M=a^E_0$.
 Moreover  the following identities hold
\begin{subequations}
\begin{align}
\sum_{k=1}^n \left(A_{ik}A_{jk}+E_{ik}E_{jk}\right)&=\delta_{ij},\quad \kappa  \in [0,1],\label{relea}\\
\sum_{k=1}^n \left(A_{ik}A_{jk}-E_{ik}E_{jk}\right)&=\delta_{ij}, \quad \kappa>1.
\end{align}
\end{subequations}
This implies that there exist some orthogonal matrices $O,O',O''$ realizing the following singular value decompositions
\cite{lupo2}:
\begin{subequations}
\begin{eqnarray}
A_{jh} &=& \sum_{j'=1}^n O_{jj'} \, \sqrt{\eta^{(n)}_{j'}} \, O'_{j'h}\, , \\
E_{jh} &=& \sum_{j'=1}^n O_{jj'} \, \sqrt{\left|\eta^{(n)}_{j'}-1\right|} \,
O''_{j'h} \,,
\end{eqnarray}
\end{subequations}
where $\eta^{(n)}_j$ are positive real numbers and the matrix $O$ is the same in both decompositions.
In terms of the following set of collective modes:
\begin{subequations}
\begin{eqnarray}
\mathrm{a'}_j &:=& \sum_{j'=1}^n O_{j'j} \, a'_{j'} \, ,  \\
\mathrm{a}_j &:=& \sum_{j'} O'_{jj'} \, a_{j'} \, , \label{a_collective}\\
\mathrm{a}^E_j &:=& \sum_{j'} O''_{jj'} \, a^E_{j'} \, ,
\end{eqnarray}
\end{subequations}
the memory channel is diagonalized into $n$ independent channels,
\begin{subequations}
\begin{align}
\mathrm{a}'_j  & = \sqrt{\eta_j^{(n)}} \,  \mathrm{a}_j -
\sqrt{1-\eta_j^{(n)}} \, \mathrm{a}^E_j \, , \quad \kappa  \in [0,1] \, , \\
\mathrm{a}'_j  & = \sqrt{\eta_j^{(n)}} \,  \mathrm{a}_j +
\sqrt{\eta_j^{(n)}-1} \, \mathrm{a}_j^{E \,\dag} \, , \quad \kappa > 1
\, .
\end{align}
\end{subequations}
In particular, if we focus on the physically relevant case in which all the modes of the environment (and the initial memory mode) are in the same thermal state with a given mean photon number $N$, the modes $\left\{ \mathrm{a}^E_j \right\}$ remain in factorized thermal states and one can conclude that the memory channel applied $n$ times is unitarily equivalent to $n$ independent memoryless attenuators or amplifiers,
\begin{equation}
\Phi_n= \mathcal E_{\eta_1^{(n)}}^{N} \otimes  \mathcal E_{\eta_2^{(n)}}^{N} \dots \otimes   \mathcal E_{\eta_n^{(n)}}^{N} .  \label{factor}
\end{equation}

An important feature of the canonical transformation \eqref{a_collective} is that annihilation operators $a_j$
are not mixed with creation operators $a_{j'}^\dag$. This means that the operation is passive, {\it i.e.}\ it does not change the total energy of the input modes and so the capacity with constrained input energy is the same for the diagonalized channel and the original one.

\subsection{Limit of infinite iterations}
In order to compute the capacity we need to take the limit infinite iterations of the memory channel.
In virtue of the previous factorization into independent channels, the capacity will depend only
on the asymptotic distribution of the gain parameters $\eta_j^{(n)}$ appearing in \eqref{factor}, in the limit of $n \rightarrow \infty$.
The set of gain parameters $\eta_j^{(n)}$ can be computed as the eigenvalues of the matrix
\begin{equation}
M^{(n)} := A A^\dag \, .
\end{equation}
The entries of the matrix $M$ can be computed from the explicit values of $A$ \cite{lupo2}, obtaining
\begin{equation}\label{theseq}
M^{(n)}_{jj'} = \delta_{jj'} + \left(\kappa_{jj'}-1\right)
\sqrt{\mu\kappa}^{|j-j'|} \, ,
\end{equation}
where
\begin{equation}
\kappa_{jj'} := \kappa + \mu(\kappa-1)^2
\sum_{h=0}^{\min{\{j,j'\}}-2} (\mu\kappa)^h \, .
\end{equation}
The asymptotic Behavior of the eigenvalues is different according to whether the combination $\mu\kappa$ is greater or lower than one.
Below threshold, {\it i.e.}\ for $\mu\kappa<1$ the sequence of matrices $M^{(n)}$ is {\it
asymptotically equivalent} \cite{toeplitz} to the (infinite)
Toeplitz matrix $M^{(\infty)}$, given by
\begin{equation}
M_{jj'}^{(\infty)} := M_{j-j'}^{(\infty)} = \delta_{jj'} - \frac{(1-\mu)(1-\kappa)}{1-\kappa\mu}\sqrt{\mu\kappa}^{|j-j'|} \, .
\end{equation}

\begin{figure*}[t]
\includegraphics[width=1 \textwidth]{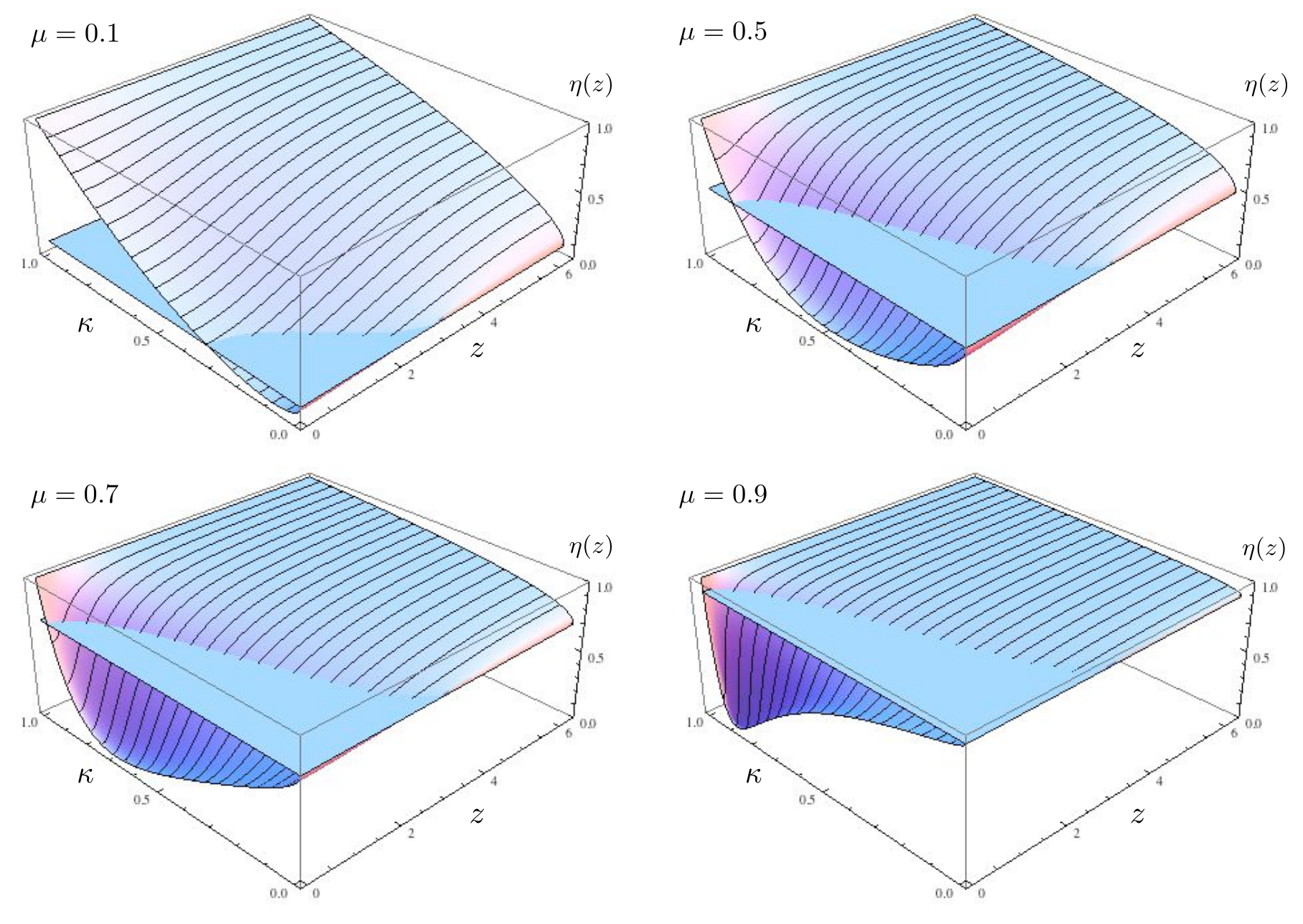}
\caption{(Color online): Asymptotic spectrum $\eta(z)$ of Eq.~(\ref{monospectrum})  for the case where ${\cal E}_{\kappa}$ of Fig.~\ref{model}
represents an attenuator
channel (i.e.  $\kappa\in[0,1]$). In this case the system is operated  below the threshold limit $\mu \kappa \leq1$ (no divergency in the spectrum occurs) and the
 values of $\eta(z)$ are always bounded below $1$ (i.e. the channels ${\cal E}_{\eta_j^{(n)}}^{N}$ entering the decomposition~(\ref{factor}) are attenuators).
In each plot the plane represents the value of $\mu$. Notice that for $\kappa = 0$ one has $\eta(z)= \mu$, while for $\kappa =1$, $\eta(z)=1$ independently from  $\eta$.
}\label{effectivetrans1}
\end{figure*}

\begin{figure*}[t]
\includegraphics[width=1 \textwidth]{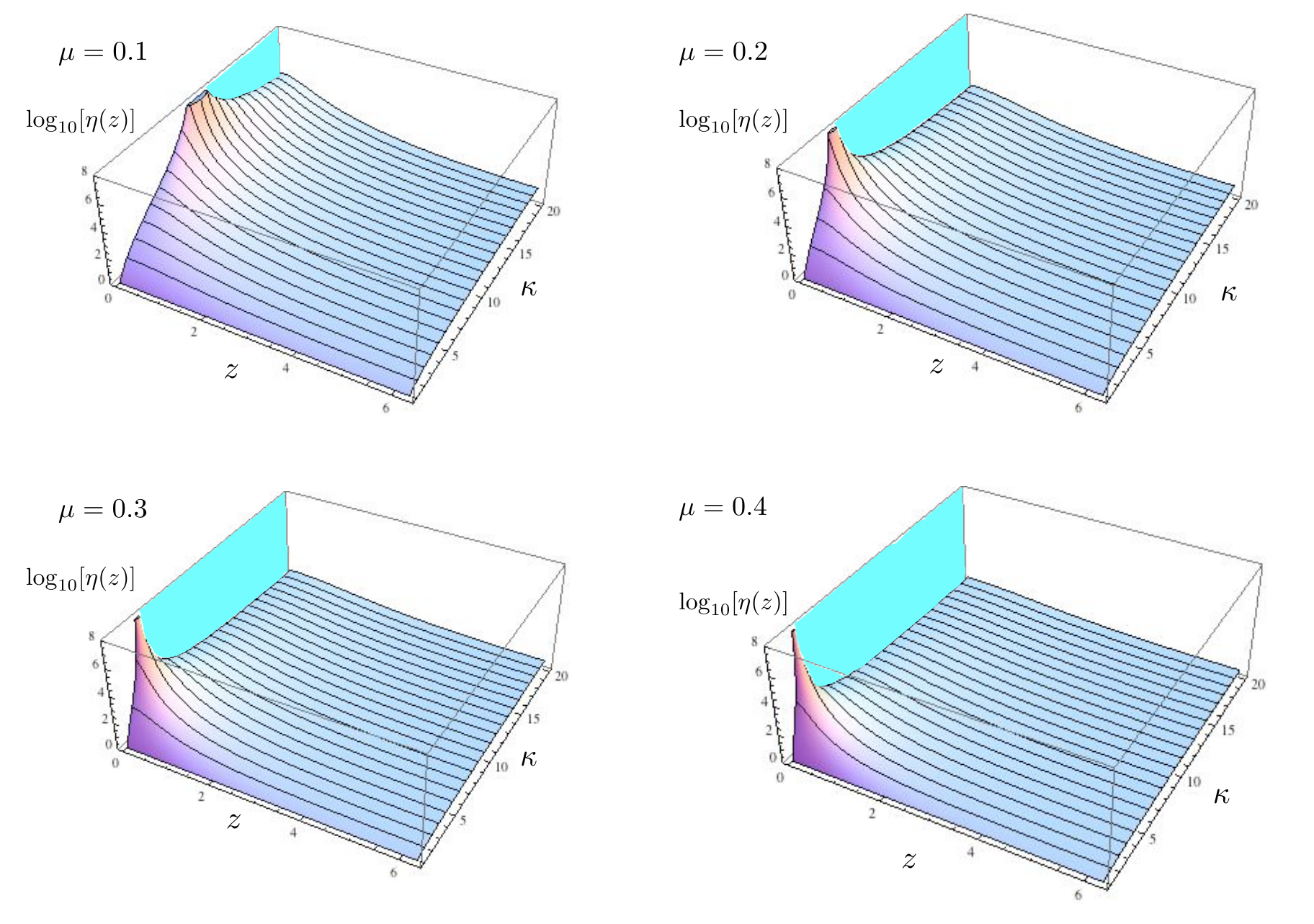}
\caption{(Color online):
Logarithm of the asymptotic spectrum $\eta(z)$ of Eq.~(\ref{monospectrum})  for the case where ${\cal E}_{\kappa}$ of Fig.~\ref{model}
represents an amplifier
channel (i.e.  $\kappa\geq 1$). In this case the
 values of $\eta(z)$ are always larger than $1$ meaning that the channels ${\cal E}_{\eta_j^{(n)}}^{N}$ entering the decomposition~(\ref{factor}) describe
 amplifiers. Above threshold (i.e. $\mu \kappa \geq 1$) the system acquires also a divergent, singular eigenvalue represented in the picture by the cyan vertical region.
}\label{effectivetrans2}
\end{figure*}

We can now exploit the full power of the Toeplitz matrices theory (see Ref.~\cite{toeplitz} for more details): the Szeg\"o theorem \cite{toeplitz} states that,
for any smooth function $F$, we have
\begin{equation}\label{szego}
\lim_{n\to\infty} \frac{1}{n} \sum_{j=1}^n F\left[\eta^{(n)}_j\right] = \int_0^{2\pi}
\frac{dz}{2\pi} F[\eta(z)] \, ,
\end{equation}
where the function $\eta(z)$ is the Fourier transform of the elements of the matrix $M^{(\infty)}$, \emph{i.e.}
\begin{equation}
\eta(z) = \sum_{j=-\infty}^\infty M^{(\infty)}_{j} e^{iz j/2} = \frac{\kappa+\mu-2\sqrt{\kappa\mu}\cos\frac{z}{2}}{1+\kappa\mu-2\sqrt{\kappa\mu}\cos\frac{z}{2}} \label{monospectrum}\, ,
\end{equation}
with $z\in[ 0, 2\pi]$ (see Fig.s~\ref{effectivetrans1}, \ref{effectivetrans2}).

Above threshold, \emph{i.e.} for $\mu\kappa>1$, the sequence of matrices
does not converge. Nonetheless, the divergence can be ascribed to a single diverging eigenvalue, and it is possible to rewrite Eq.~\eqref{theseq} as the sum of two terms:
\begin{equation}\label{commuting}
M^{(n)} = c^{(n)} P^{(n)} + \Delta{M}^{(n)} \, ,
\end{equation}
where the $P^{(n)}$ are rank one projectors, $c^{(n)}$ is a
diverging sequence of positive real numbers, and $\Delta{M}^{(n)}$
is a sequence of matrices which asymptotically converges towards the
(infinite) Toeplitz matrix $\Delta{M}^{(\infty)}$, given by
\begin{equation}\label{Mrescaled}
\Delta{M}^{(\infty)}_{jj'} = \delta_{jj'}+\frac{(1-\mu)(\kappa-1)}{\mu\kappa-1}
\frac{1}{\sqrt{ \mu\kappa }^{|j-j'|}} \, .
\end{equation}
(See the appendix of \cite{lupo2} for the expressions of $P^{(n)}$ and $\Delta{M}^{(n)}$). It is possible to prove that for $n\rightarrow
\infty$, the matrices $P^{(n)}$  and  $\Delta{M}^{(n)}$  commute, and we can conclude that, as promised,
 the spectrum of the matrix \eqref{theseq} is
asymptotically composed of only one diverging eigenvalue [corresponding to the
diverging sequence $c^{(n)}$] and of the asymptotic spectrum of the infinite Toeplitz matrix \eqref{Mrescaled}.
As for the below threshold case, the latter is given by the Fourier transform of the matrix elements, where the Fourier transform $\eta(z)$ is given by Eq.\ \eqref{monospectrum} analytically continued to the region $\mu\kappa>1$.

Finally it remains to consider the case $\mu\kappa=1$. At this threshold,  the matrix $M^{(n)}$ can be
expressed as
\begin{equation}\label{theseqTHR}
M^{(n)}_{jj'} = \delta_{jj'} + (1-\mu) + \frac{(1-\mu)^2}{\mu} \min\{j,j'\} \;.
\end{equation}
In this case it appears not feasible to extract the asymptotic spectrum. From a practical point of view however this is not a real problem since any real physical channel will always fall into one of the two classes characterized by $\mu \kappa >1 $ or $\mu \kappa < 1$, respectively.

It is important to stress that for any $\mu\in[0,1]$, in the thermal attenuator case ($\kappa\in[0,1]$) all the channels in the asymptotic diagonal decomposition \eqref{factor} are also thermal attenuators, \emph{i.e.} $\eta(z)\in[0,1]$ for any $z\in[0,2\pi]$. The same happens in the amplifier case, \emph{i.e.} if $\kappa>1$ also $\eta(z)>1$ for any $z\in[0,2\pi]$.

\section{Capacities}\label{sec:capac}
In this section we will compute the capacity of the memory channel model of the previous section, with the environment in a thermal multi-mode state with fixed temperature and associated mean photon number per mode $N$.

Let $\Phi_P$ be the mapping describing the input-output relations of the first $P$-channel uses of the model depicted in Fig.~\ref{model}.
Since any input influences all the following outputs, its classical capacity cannot be directly computed as in Eq.~(\ref{capacityformula}).
Still, thanks to the fact that
$\Phi_P$ can be expressed as a tensor product of $P\gg 1$ independent maps
of effective transmissivities $\eta^{(P)}_j$ (see Eq.~(\ref{factor})), a close formula for $C$ can be derived.
The fundamental observation here is that, even though in general the  $\eta^{(P)}_j$ will differ from each other,
 for large enough $P$ one can organize them into subgroups each containing a number of elements of order $P$, and characterized by an almost identical
 value of the transmissivity distributed according to the continuous function~$\eta(z)$ of Eq.~(\ref{monospectrum}).
Consider next the channel $\Phi_{2P}$. Its effective transmissivities are different, but they are taken from almost the same distribution, therefore we can write
\begin{equation}
\Phi_{2P}\simeq \Phi_P\otimes\Phi_P\;.
\end{equation}
Iterating, we get
\begin{equation}
\Phi_{\ell P}\simeq\Phi_P^{\otimes \ell}\;,
\end{equation}
and we have managed to express $\Phi_n$ for $n\to\infty$ as the limit of infinite uses of a fixed memoryless channel.

Let's formalize this procedure: we fix $P\gg1$, and take $n=\ell P$. We label the eigenvalues $\eta^{(n)}_j$ in increasing order ($\eta^{(n)}_j\leqslant\eta^{(n)}_{j'}$ if $j<j'$), and divide them into $P$ groups, the $p$th one being made by $\left\{\left.\eta^{(n)}_j\right|(p-1)\ell < j \leqslant
p\ell \right\}$.
Let $\underline{\eta}^{(P)}_p$ and $\overline{\eta}^{(P)}_p$ be respectively the infimum and the supremum of the $p$th group over all $\ell$:
\begin{subequations}
\begin{align}
\underline{\eta}^{(P)}_p = & \inf_\ell \inf_{(p-1)\ell < j \leqslant
p\ell} \eta^{(\ell P)}_j \, , \\
\overline{\eta}^{(P)}_p = & \sup_\ell \sup_{(p-1)\ell < j \leqslant
p\ell} \eta^{(\ell P)}_j \, .
\end{align}
\end{subequations}
Now, the two collections of transmissivities $\underline{\eta}^{(P)}_p$ and $\overline{\eta}^{(P)}_p$ identify two memoryless $P$-mode gaussian channels.
Let $\phi(\eta,N)$ be the Gaussian attenuator / amplifier with transmissivity $\eta\geqslant0$, mixing the input with a thermal state with mean photon number $N$. Remembering that $\phi(\eta,N)\phi(\eta',N)=\phi(\eta\eta',N)$ and that the capacity decreases under composition of channels, if we replace each transmissivity with the supremum or the infimum of its group, the capacity will increase or decrease, respectively.
Each group has exactly $\ell$ eigenvalues, so the $n$ uses of the single mode memory channel can be compared to $\ell$ uses of these two $P$-mode channels, and letting $\ell\to\infty$ we can bound the capacity with
\begin{equation}\label{IMPOC}
\underline{C}^{(P)} \leqslant C \leqslant \overline{C}^{(P)} \, ,
\end{equation}
where $\underline{C}^{(P)}$ and $\overline{C}^{(P)}$ are precisely the capacities of these $P$-mode channels with transmissivities $\left\{ \underline{\eta}^{(P)}_p \right\}$, $\left\{ \overline{\eta}^{(P)}_p \right\}$.
As customary, to keep them finite we impose a constraint on the input mean energy:
\begin{equation}\label{constraint}
\frac{1}{n} \sum_{j=1}^n \Tr\left[ \rho^{(n)} a^\dag_j a_j \right]
\leqslant E \, ,
\end{equation}
where $n$ is the number of uses of the channel and $\rho^{(n)}$ is the joint input density matrix.
As already stressed, this constraint looks identically if expressed in terms of the collective modes \eqref{a_collective}, since they are related to the original ones by an orthogonal matrix.
\subsection{Thermal attenuator}
Let's first consider the case of the attenuating thermal memory channel, \emph{i.e.} $\kappa\leq1$.
It has recently been proven \cite{conj1} that the $\chi$ capacity of successive uses of Gaussian phase-insensitive channels is additive also if they are different:
\begin{equation}
\chi(\Phi_1\otimes\ldots\otimes\Phi_n)=\chi(\Phi_1)+\ldots+\chi(\Phi_n)\;.
\end{equation}
Then the capacity of our two $P$-mode channels can be simply obtained by summing \eqref{capbs} over all modes, yielding the bounds
\begin{subequations}
\label{boundC}
\begin{align}
\underline{C}^{(P)} & = \frac{1}{P} \sum_{p=1}^P & \left( g\left[\underline{\eta}^{(P)}_p \underline{N}_p+\left(1-\underline{\eta}^{(P)}_p\right)N_T\right]+\right.\nonumber\\
&&\left.-g\left[\left(1-\underline{\eta}^{(P)}_p\right)N_T\right]\right) \, ,\\
\overline{C}^{(P)}  & = \frac{1}{P} \sum_{p=1}^P&
\left(g\left[\overline{\eta}^{(P)}_p \overline{N}_p+\left(1-\overline{\eta}^{(P)}_p\right)N_T\right]+\right.\nonumber\\
&&\left.-g\left[\left(1-\overline{\eta}^{(P)}_p\right)N_T\right]\right) \, ,
\end{align}
\end{subequations}
where
\begin{equation}
g(x)=(x+1)\ln(x+1)-x\ln x\;,
\end{equation}
and the parameters $\underline{N}_p$, $\overline{N}_p$ describe the optimal distribution of the mean photon number of the modes and must satisfy the constraints
\begin{subequations}
\begin{align}
&\underline{N}_p,\,\overline{N}_p\geqslant\label{positive}
0\\
&\frac{1}{P}\sum_{p=1}^P \underline{N}_p=\frac{1}{P}\sum_{p=1}^P \overline{N}_p=E\;.
\end{align}
\end{subequations}
If the positivity constraint \eqref{positive} were not there, these optimal values could be computed with the Lagrange multiplier method, yielding
\begin{equation}\label{Lagrange}
\underline{N}_p = \frac{1}{\underline{\eta}^{(P)}_p}\left(\frac{1}{e^{\underline{\lambda}/\underline{\eta}^{(P)}_p} - 1}-\left(1-\underline{\eta}^{(P)}_p\right)N\right)\;,
\end{equation}
and the analogue for $\overline{N}_p$.
Taking the limit $P\to\infty$ and applying \eqref{szego}, the two bounds converge to the same quantity and we get
\begin{align}\label{classical}
C = \int_0^{2\pi} \frac{dz}{2\pi}& \left(g\left[\eta(z) N(z)+\left(1-\eta(z)\right)N\right]+\right.\nonumber\\
&\left.-g\left[\left(1-\eta(z)\right)N\right]\right) \quad \kappa\in[0,1] \, .
\end{align}
In the zero temperature case $N=0$ the expression \eqref{Lagrange} is positive definite. As $N$ grows, \eqref{Lagrange} is no more guaranteed to be positive, and we have to impose this constraint by hand. Then, above a certain critical temperature the optimal energy distribution $N(z)$ will vanish for $0\leqslant z\leqslant z_0$. Physically, this means that it is convenient to concentrate all the energy on a fraction $\frac{2\pi-z_0}{2\pi}$ of all the beam-splitters. We will show in section \ref{sec:lagr} that to determine the optimal energy distribution we can still use the Lagrange multipliers, with the only caveat that $N(z)$ is given now by the positive part of what we would have got without the energy constraint:
\begin{equation}
N(z)=\frac{1}{\eta(z)}\left(\frac{1}{e^\frac{\lambda}{\eta(z)}-1}-(1-\eta(z))N\right)^+\;,\label{nzbs}
\end{equation}
where $f^+(z)=[ f(z) + |f(z)| ]/2$ is the positive part of $f$. The energy constraint reads as expected
\begin{equation}
\int_0^{2\pi} \frac{dz}{2\pi} N(z) = E \, .\label{enconstr}
\end{equation}
We notice that the function $\eta$ is symmetric in $\mu$ and $\kappa$, \emph{i.e.}
\begin{equation}
\eta(\mu,\kappa,z)=\eta(\mu'=\kappa,\;\kappa'=\mu,\;z)\;.
\end{equation}
Since $\mu$ and $\kappa$ appear in the computation of the capacity only through $\eta$, the channel with parameters $(\mu',\kappa')$ has the same capacity of the original one, \emph{i.e.} we can exchange the memory with the transmissivity. Then, varying the memory with fixed transmissivity has the same effect on the capacity as varying the transmissivity for fixed memory. In Fig. \ref{capacityfig} we report the capacity of the channel as a function of the temperature.

\subsection{Thermal amplifier}
The minimum output entropy conjecture lets us compute the capacity also in the amplifier case $\kappa>1$. Now, all the transmissivities are greater than 1, so the capacity decreases as they increase and the two bounds \eqref{boundC} are inverted:
\begin{subequations}
\begin{align}
\overline{C}^{(P)} & = \frac{1}{P} \sum_{p=1}^P&\left( g\left[\underline{\eta}^{(P)}_p \underline{N}_p+\left(\underline{\eta}^{(P)}_p-1\right)\left(N+1\right)\right]+\right.\nonumber\\
&&\left.-g\left[\left(\underline{\eta}^{(P)}_p-1\right)\left(N+1\right)\right]\right) \, ,\\
\underline{C}^{(P)}  & = \frac{1}{P} \sum_{p=1}^P&
\left(g\left[\overline{\eta}^{(P)}_p \overline{N}_p+\left(\overline{\eta}^{(P)}_p-1\right)\left(N+1\right)\right]+\right.\nonumber\\
&&\left.-g\left[\left(\overline{\eta}^{(P)}_p-1\right)\left(N+1\right)\right]\right) \, .
\end{align}
\end{subequations}
As in the thermal attenuator case, we take the limit $P\to\infty$. Above the threshold ($\mu\kappa>1$) one of the eigenvalues is diverging but, being only one, it does not contribute in the limit, so the capacity is still fully determined by the infinite Toeplitz matrix $\Delta M^{(\infty)}$ yielding
\begin{align}
C = \int_0^{2\pi} \frac{dz}{2\pi} &\left(g\left[\eta(z) N(z)+\left(\eta(z)-1\right)\left(N+1\right)\right]\right.+\nonumber\\
&\left.-g\left[\left(\eta(z)-1\right)\left(N+1\right)\right]\right) \label{classicalamp}\, ,
\end{align}
where as before $N(z)$ is determined by the Lagrange multiplier method, with the caveat of taking the positive part of the resulting function
\begin{equation}
N(z)=\frac{1}{\eta(z)}\left(\frac{1}{e^\frac{\lambda}{\eta(z)}-1}-(\eta(z)-1)(N+1)\right)^+\;,\label{nzamp}
\end{equation}
and with the same constraint on the mean energy
\begin{equation}
\int_0^{2\pi} \frac{dz}{2\pi} N(z) = E \, .
\end{equation}
We notice that in \eqref{nzamp} the positive part is at least in principle necessary also in the case of zero temperature.

Also the amplifier enjoys a sort of duality between $\kappa$ and $\mu$: the function $\eta(\mu,\kappa,z)$ satisfies
\begin{equation}
\eta(\mu,\kappa,z)=\eta\left(\mu'=\frac{1}{\kappa},\;\kappa'=\frac{1}{\mu},\;z\right)\;.
\end{equation}
Noticing that $\kappa'\mu'=\frac{1}{\kappa\mu}$, this relation associates to any channel identified by $(\mu,\kappa)$ above threshold ($\mu\kappa>1$) the new one identified by $(\mu',\kappa')$, which is below threshold.
Then, to investigate the capacity regions as function of the parameters, it is sufficient to consider only the channels below threshold.

\begin{figure}[t]
\centering
  \includegraphics[width=\columnwidth]{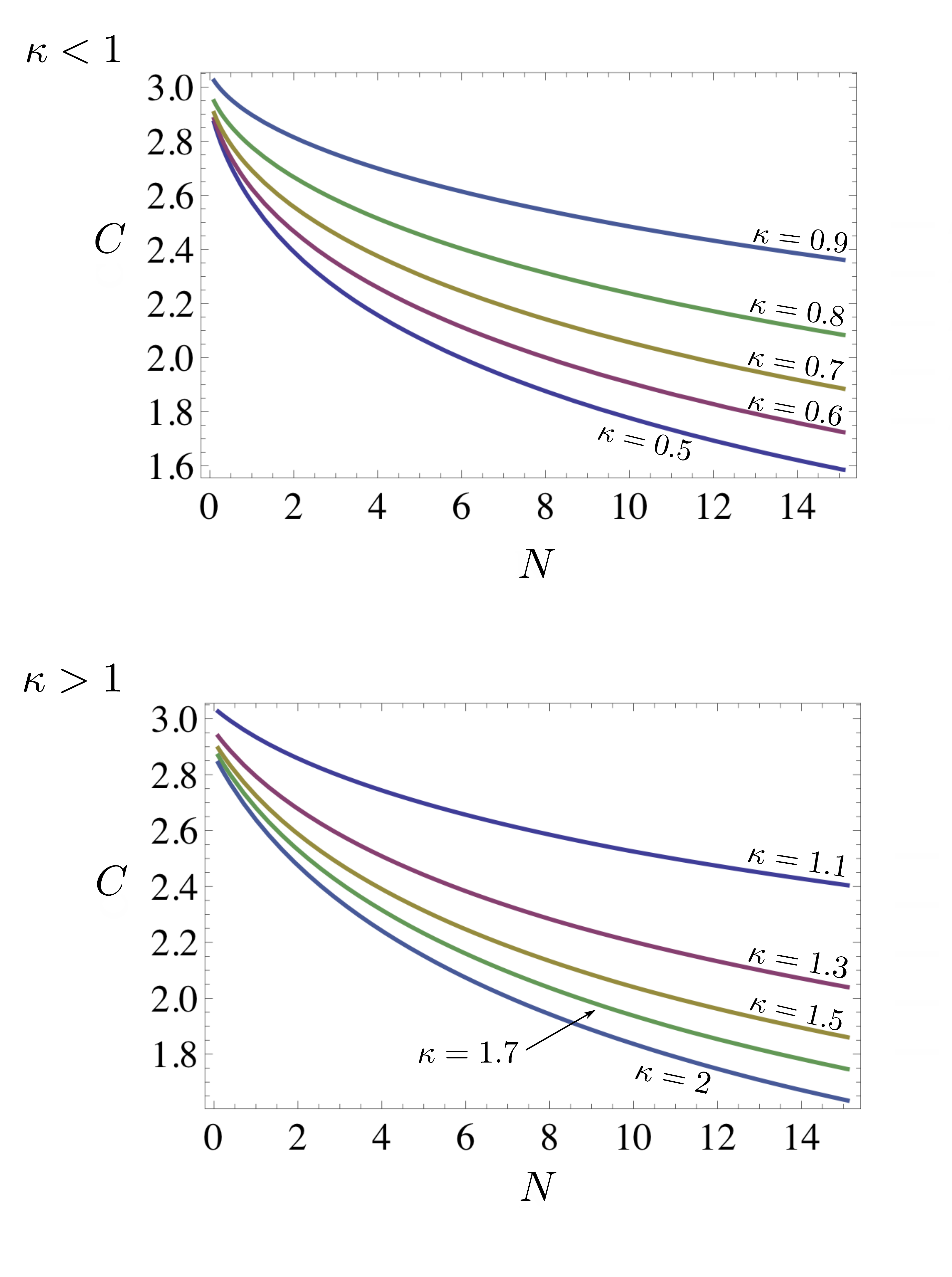}\\
  \caption{(Color online): Capacity (in nats / channel use) as a function of the thermal photon number $N$ for $\mu=0.8$ and mean input energy $E=8$ for various values of the transmissivity $\kappa$.  In particular the upper panel refers to the case where the map ${\cal E}_{\kappa}$ of Fig.~\ref{model}  is  an attenuator (i.e. $\kappa \in [0,1]$), while the lower panel to the case where  ${\cal E}_{\kappa}$ is an amplifier ($\kappa \geq 1$).    As expected, the capacity is degraded by the temperature and enhanced if the transmissivity is close to unity.}
\label{capacityfig}
\end{figure}

In Fig. \ref{capacityfig} we report the capacity of the thermal memory channel as a function of the thermal photon number $N$. As for the thermal attenuator, the capacity is degraded by the temperature and enhanced by the memory.

\subsection{Optimal encoding and decoding}
We have seen how the optimal encoding is a coherent-state one with Gaussian weights in the normal mode decomposition $\{\mathrm{a}_j\}$ introduced in~Eq.~(\ref{a_collective})  in which the channel is diagonal.
They  are related to the input modes $\{ a_j \}$ by a passive orthogonal transformation, and since such transformations send coherent states to coherent states, the latter are also not entangled. However, since the optimal coding requires a non-uniform energy distribution among the $\{\mathrm{a}_j\}$, the  modes $\{ a_j \}$ will be classically correlated. Then this optimal coding can be achieved by independent uses of the channel, but the probabilities of choosing a particular coherent state will be correlated among the various inputs.

Since also in the case of multiple uses of a fixed memoryless channel the optimal decoding requires measures entangled among the various outputs \cite{holevo}, in our case the preprocessing with an orthogonal passive transformation to convert the physical basis into the diagonal one does not add further complications to the procedure.

Above threshold ($\mu\kappa>1$), the diverging eigenvalue signals the presence of an input mode that gets amplified by a factor which increases indefinitely with the number of channel uses. Then, even if such mode is left in the vacuum, the corresponding output mode will have a very high energy, and could in principle lead the beam-splitter used in the decoding procedure to a nonlinear regime. The experimentally achievable capacity could then be lower than the theoretical bound, depending on the stability of the decoding device when dealing with high energy inputs.

\subsection{Trivial cases}
There are some particular values of the parameters for which the capacity can be computed analytically.
\begin{itemize}
\item $\kappa=1$ or $\mu=1$

This case corresponds to the identity channel ($\kappa=1$) or to the perfect memory channel ($\mu=1$). In both cases, $\eta(z)=1$ and the capacity is the one of the identity channel with mean energy $E$:
\begin{equation}
C=g(E)\;.
\end{equation}
An intuitive explanation of the result for the perfect memory channel can be given: since $\mu=1$, the first $n$ output modes $\left\{a'_i\right\}$ are a linear combination only of the first $n$ input modes $\left\{a_i\right\}$ and the first memory mode $a_1^M$, and the environment modes $\left\{a_i^E\right\}$ do not play any role. Now we can imagine that in the large $n$ limit the mode $a_1^M$ is no more relevant, and the channel behaves almost as if the output modes were an invertible linear combination of the input ones. This combination can be inverted in the decoding, recovering (almost) the identity channel.

\item $\kappa\to\infty$

This is the case of infinite amplification. Here $\eta(z)=\frac{1}{\mu}$, and the capacity is the one of the amplifier with amplification factor $\frac{1}{\mu}$
\begin{equation}
C=g\left(\frac{E}{\mu}+\frac{1-\mu}{\mu}\left(N+1\right)\right)-g\left(\frac{1-\mu}{\mu}\left(N+1\right)\right)\;.
\end{equation}

\item $\kappa=0$

This is the case of infinite attenuation, in which all the signal is provided by the memory. Here the $n$-th input mode $a_n$ does not influence at all the $n$-th output $a'_n$, but it directly mixes with the $n+1$-th environmental mode $a_{n+1}^E$ through the beam-splitter with transmissivity $\mu$ to give the $n+1$-th output $a'_{n+1}$. Then the only memory effect is a translation of the inputs, and the channel behaves as a thermal attenuator with transmissivity $\mu$. Indeed, as shown in Fig. \ref{effectivetrans1}, here $\eta(z)=\mu$, and the capacity matches the attenuator one \cite{CC}:
\begin{equation}
C=g\left(\mu E+(1-\mu)N\right)-g\left((1-\mu)N\right)\;.
\end{equation}

\item $\mu=0$

This is the memoryless case, and the capacity is the one of the thermal attenuator / amplifier with transmissivity $\kappa$:
\begin{subequations}
\begin{align}
C&=g\left(\kappa E+(1-\kappa)N\right)-g\left((1-\kappa)N\right)\;,\\
C&=g\left(\kappa E+(\kappa-1)\left(N+1\right)\right)-g\left((\kappa-1)\left(N+1\right)\right)\;.
\end{align}
\end{subequations}
\end{itemize}

\subsection{Additive noise channel}
The one--mode additive noise channel adds to the covariance matrix $\sigma$ of the input state a multiple of the identity:
\begin{equation}
\sigma\mapsto\sigma+N_C\mathbbm{1}\;.
\end{equation}
A beam-splitter of transmissivity $\eta$, mixing the input with a thermal state with mean photon number $N$, performs instead a convex combination of the corresponding covariance matrices:
\begin{equation}\label{bscm}
\sigma\mapsto\eta\sigma+(1-\eta)\left(N+\frac{1}{2}\right)\mathbbm{1}\;.
\end{equation}
The additive noise channel can now be recovered in the limit $\eta\to1^-$ with the second addend of \eqref{bscm} kept fixed, i.e. with
\begin{equation}
(1-\eta)\left(N+\frac{1}{2}\right)=N_C\;,\qquad\eta\to1^-\;,\qquad N\to\infty\;.
\end{equation}
It is then natural to consider what happens to our model for the memory channel in the limit $N\to\infty$, $\kappa\to1^-$ with fixed $(1-\kappa)\left(N+\frac{1}{2}\right)=N_C$. We start from the expression \eqref{outputmodes} which expresses the output modes in terms of the input and the (thermal) environment. From the expressions for the matrices $A$ and $E$ in \cite{lupo2} it is easy to show that, since they do not depend on $N$, their limit for $\kappa\to1$ are $A\to\mathbbm{1}$ and $E\to0$, respectively. Physically, this happens because for $\kappa=1$ the channel is the identity and the output is equal to the input. We will now compute the expectation values of all the operators quadratic in the output modes, \emph{i.e.} the output covariance matrix.
We remember that, since the input and the environment are in a completely factorized state,
\begin{subequations}
\begin{eqnarray}
&\left\langle a_i a_j^E\right\rangle=\left\langle a_i^\dag a_j^E\right\rangle=\left\langle a^E_i a_j^E\right\rangle=0\;,\\
&\left\langle {a^E_i}^\dag a_j^E\right\rangle=N\delta_{ij}\;.
\end{eqnarray}
\end{subequations}
We have then
\begin{subequations}
\begin{eqnarray}
\left\langle a_i' a_j'\right\rangle&=&\left\langle a_i a_j\right\rangle\;,\\
\left\langle {a_i'}^\dag a_j'\right\rangle&=&\left\langle {a_i}^\dag a_j\right\rangle+\lim_{N\to\infty}N\sum_k E_{ik}E_{jk}\;,
\end{eqnarray}
\end{subequations}
where the limit is nontrivial since the matrix $E$ depends on $\kappa$, which changes with $N$. Recalling \eqref{relea}
\begin{equation}
AA^T+EE^T=\mathbbm{1}\;,
\end{equation}
and from the expression for $AA^T=AA^\dag$ in \cite{lupo2} it is easy to prove that
\begin{equation}
\lim_{N\to\infty}N\sum_k E_{ik}E_{jk}=N_C\mu^\frac{|i-j|}{2}\;,
\end{equation}
so
\begin{equation}
\left\langle {a_i'}^\dag a_j'\right\rangle=\left\langle {a_i}^\dag a_j\right\rangle+N_C\mu^\frac{|i-j|}{2}\label{additivenoise}\;.
\end{equation}
If we look only at a single output mode $a_i'$, throwing away all the others, \eqref{additivenoise} becomes
\begin{equation}
\left\langle {a_i'}^\dag a_i'\right\rangle=\left\langle {a_i}^\dag a_i\right\rangle+N_C\;,
\end{equation}
\emph{i.e.} the reduced channel exactly adds classical noise $N_C$.
However, for nonzero memory ($\mu>0$), $N_C\mu^\frac{|i-j|}{2}$ is nonzero also for $i\neq j$: the added noise is correlated among the various outputs, and the resulting channel is not simply the product of $n$ independent additive noise ones.
We expect this correlation to enhance the capacity: looking at the limit of our formula \eqref{classical}, we will see that it is effectively so. Let's look at this limit in the normal modes variables. Remembering that the environment associated to the operators $\mathrm{a}_j^E$ is still in a factorized thermal state with temperature $N$, we have
\begin{subequations}
\begin{eqnarray}
\left\langle \mathrm{a}_i' \mathrm{a}_j'\right\rangle&=&\left\langle \mathrm{a}_i \mathrm{a}_j\right\rangle\;,\\
\left\langle {\mathrm{a}_i'}^\dag \mathrm{a}_j'\right\rangle&=&\left\langle {\mathrm{a}_i}^\dag \mathrm{a}_j\right\rangle+\delta_{ij}\lim_{N\to\infty}N\left(1-\eta^{(n)}_i\right)\;,
\end{eqnarray}
\end{subequations}
and since
\begin{equation}
\lim_{N\to\infty}(1-\eta(z))N=\frac{N_C(1-\mu)}{1+\mu-2\sqrt{\mu}\cos\frac{z}{2}}\;,\label{distraddnoise}
\end{equation}
in the limit of infinite channel uses we get a factorized additive noise channel, but with the added noise depending on the mode and distributed according to \eqref{distraddnoise}. This model for an additive noise channel with memory coincides with the one considered in \cite{additiven1,CERFA}, derived starting from correlated translations with Gaussian weights.

First, notice that $\eta(z)$ does not depend on $N$, and $\lim_{\kappa\to1}\eta(z)=1$.
Let's compute the limit of the expression for $N(z)$ \eqref{nzbs}:
\begin{equation}
N(z)=\left(\frac{1}{e^\lambda-1}-\lim_{N\to\infty}(1-\eta(z))N\right)^+\;.
\end{equation}
From the expression for $\eta(z)$ \eqref{monospectrum} we can compute the limit

so that
\begin{equation}
N(z)=\left(\frac{1}{e^\lambda-1}-\frac{N_C(1-\mu)}{1+\mu-2\sqrt{\mu}\cos\frac{z}{2}}\right)^+\label{naddit}\;.
\end{equation}
For simplicity, we consider only the case in which the positive part in \eqref{naddit} is not needed.
The mean energy constraint \eqref{enconstr} becomes
\begin{equation}
\frac{1}{e^\lambda-1}=N_C+E\;,
\end{equation}
where we have used that
\begin{equation}
\int_0^{2\pi}\frac{1-\mu}{1+\mu-2\sqrt{\mu}\cos\frac{z}{2}}\frac{dz}{2\pi}=1\;,
\end{equation}
and we have for the positivity constraint on $N(z)$
\begin{equation}
E\geq\frac{2N_C\sqrt{\mu}}{1-\sqrt{\mu}}\;.
\end{equation}
Finally, we can compute the capacity taking the limit of \eqref{classical}:
\begin{equation}
C=g(E+N_C)-\int_0^{2\pi}g\left(\frac{N_C(1-\mu)}{1+\mu-2\sqrt{\mu}\cos\frac{z}{2}}\right)\;.\label{capaddn}
\end{equation}
Since $g(x)$ is concave, the LHS of \eqref{capaddn} decreases if we take the integral inside $g$, so
\begin{equation}
C\geq g(E+N_C)-g(N_C)\;.  \label{boundC2}
\end{equation}
The right-hand-side  of \eqref{boundC2} is exactly the capacity of the single mode additive noise channel, \emph{i.e.} the correlation of the added noise enhances the capacity as expected.

\section{Optimal energy distribution}\label{sec:lagr}
In this section we will prove that the Lagrange multipliers method with the caveat of taking the positive part in \eqref{nzbs} and \eqref{nzamp} works also with the positivity constraint \eqref{positive}, and we will analyze the resulting optimal energy distribution $N(z)$.

\subsection{The proof}
The function $\eta(z)$ is increasing for the thermal attenuator ($\kappa<1$) and decreasing for the amplifier ($\kappa>1$), \emph{i.e.} the channel with transmissivity $\eta(z)$ always improves as $z$ increases.
For simplicity here we consider only the thermal attenuator case, the amplifier one being completely analogous.

Let $\widetilde{N}(z,w)$ be the Lagrange multipliers solution in the interval $w\leqslant z\leqslant2\pi$ which maximizes the capacity
\begin{align}
C = \int_w^{2\pi} \frac{dz}{2\pi} &\left(g\left[\eta(z) \widetilde{N}(z,w)+\left(\eta(z)-1\right)\left(N+1\right)\right]\right.+\nonumber\\
&\left.-g\left[\left(\eta(z)-1\right)\left(N+1\right)\right]\right)\label{capw}
\end{align}
with the mean energy constraint
\begin{equation}
\int_{w}^{2\pi}\frac{dz}{2\pi}\widetilde{N}(z,w)dz=E\;,\label{constrn}
\end{equation}
where the integrals are restricted to $w\leqslant z\leqslant2\pi$ and we do not care about the positivity of $N(z,w)$.
Such solution is given by
\begin{equation}
\widetilde{N}(z,w)=\frac{1}{\eta(z)}\left(\frac{1}{e^\frac{\lambda}{\eta(z)}-1}-(1-\eta(z))N\right),
\end{equation}
where the multiplier $\lambda$ is determined by the constraint \eqref{constrn} (strictly speaking, with $\widetilde{N}(z,w)$ we mean the function analytically continued to the whole interval $0\leqslant z\leqslant2\pi$).

Let $N(z)$ be the optimal positive distribution of the photons. Since it is better to use more energy in the better channels, $N(z)$ must be increasing: if not, we could move a bit of energy from a bad channel to a better one with less energy, and this would increase the capacity.
Let $N(z)$ be zero for $0\leqslant z<z_0$, and strictly positive for $z_0<z\leqslant2\pi$.
In particular $N(z)$ is the optimal solution among all the functions equal to zero for $0\leqslant z<z_0$ and strictly positive for $z_0<z\leqslant2\pi$.
We consider all the infinitesimal variations $N(z)+\delta N(z)$ satisfying the mean energy constraint and such that $\delta N(z)$ is nonzero only in the interval $z_0<z\leqslant2\pi$. Since $N(z)$ is strictly positive there, $N(z)+\delta N(z)$ is still positive for infinitesimal $\delta N$, so it is a legal positive photon distribution. For its optimality $N(z)$ must be a stationary point of the capacity for all such variations, but this means exactly that $N(z)$ is the solution of the Lagrange multipliers method $\widetilde{N}(z,z_0)$:
\begin{equation}
N(z)=\widetilde{N}(z,z_0)\theta(z-z_0)\;,
\end{equation}
where $\theta(z)$ is the step function.

We now claim that $\widetilde{N}(z_0,z_0)$ must be zero. Let's suppose $\widetilde{N}(z_0,z_0)>0$.
Since $\widetilde{N}(z,w)$ is continuous in $w$, we can choose a $w_0<z_0$ such that $\widetilde{N}(z,w_0)$ is strictly positive in the whole interval $w_0<z\leqslant2\pi$. Then, $\widetilde{N}(z,w_0)\theta(z-w_0)$ is an admissible solution. Since also $N(z)$ has been considered in the maximization problem \eqref{capw} defining $\widetilde{N}(z,w_0)$, the latter must achieve a greater capacity than the former, impossible.

For the same argument used with $N(z)$, $\widetilde{N}(z,z_0)$ must be increasing within each interval where it is positive, and since it is continuous in $z$ it must be negative for $0\leqslant z<z_0$ and positive for $z_0<z\leqslant2\pi$. Then we can finally write as promised $N(z)$ as
\begin{equation}
N(z)=\frac{1}{\eta(z)}\left(\frac{1}{e^\frac{\lambda}{\eta(z)}-1}-(1-\eta(z))N\right)^+,\,z\in[0,2\pi],\label{nz}
\end{equation}
where $f^+(z)$ is the positive part of $f$.

\subsection{Analysis of the optimal distribution}
The typical behavior of $N(z)$ in the attenuator case is shown in Fig. \ref{n(z)}. It is increasing, as it has to be. We can identify a critical temperature $N_{crit}$, that for our choice of the parameters ($\kappa=0.9$, $\mu=0.8$, $E=8$) is nearly $N_{crit}\sim0.8$. Below this critical value, $N(z)$ approaches a constant positive value for $z\to0$, \emph{i.e.} the optimal configuration exploits all the beam-splitters. Above the critical value, $N(z)$ is zero on a finite interval $[0,z_0]$, \emph{i.e.} the optimal configuration does not use at all a finite fraction $\frac{z_0}{2\pi}$ of the beam-splitters, being more convenient to concentrate all the energy on the other ones.

The behavior of $N(z)$ in the amplifier case is shown in Fig. \ref{n(z)amp}. It is completely analogous to the thermal attenuator, but for our choice of the parameters ($\kappa=1.1$, $\mu=0.8$, $E=8$) the critical temperature is much greater, $N_{crit}\sim 9.8$.

An analysis of the fraction $\frac{z_0}{2\pi}$ (remember that $z_0$ ranges from $0$ to $2\pi$) of the unused beam-splitters is presented in Fig. \ref{fig:Z0}. For fixed $\kappa$ and $\mu$, for zero temperature ($N=0$) all the beam-splitters are exploited and $z_0=0$; then $z_0$ remains zero up to the critical temperature $N_{crit}$, and grows for $N>N_{crit}$.
We can notice that for typical parameters, the critical value $N_{crit}$ for the beam-splitter is much lower than for the amplifier.

We will now show that in the infinite temperature limit ($N\to\infty$), $z_0$ tends to $2\pi$, and the optimal configuration concentrates all the energy on an infinitesimal fraction of the beam-splitters. First, notice that for $N\to\infty$ the multiplier $\lambda$ in \eqref{nz} must tend to zero, and we can approximate $e^{\lambda/\eta}-1\sim\lambda/\eta$, getting
\begin{equation}
N(z)=\left(\frac{1}{\lambda}-\left(\frac{1}{\eta(z)}-1\right)N\right)\theta(z-z_0)+\mathcal{O}(1)\;,
\end{equation}
where $z_0$ is the point where $N(z)$ vanishes, given by
\begin{equation}
\frac{1}{\lambda}=\left(\frac{1}{\eta(z_0)}-1\right)N\;.
\end{equation}
The energy constraint \eqref{enconstr} can be now written as
\begin{equation}
E=N\int_{z_0}^{2\pi}\left(\frac{1}{\eta(z_0)}-\frac{1}{\eta(z)}\right)\frac{dz}{2\pi}+\mathcal{O}(1)\;,
\end{equation}
and since $\eta(z)$ is strictly increasing, the only way to keep $E$ finite for $N\to\infty$ is to let $z_0\to2\pi$, \emph{i.e.} in the high temperature limit all the energy is concentrated on an infinitesimal fraction of the beam-splitters.

The minimum energy $E_{crit}$ for which all the beam-splitters are exploited is shown in Fig. \ref{fig:Encrit} for various values of the temperature $N$.
We know that for $\kappa=0,1$ and $\kappa\to\infty$ no beam-splitter is left unused, and indeed $E_{crit}=0$ at these points.
As expected, $E_{crit}$ always grows with the temperature.
In the attenuator case, we notice a divergence of $E_{crit}$ for $\kappa=\mu$ ($\mu=0.8$ in the plot). Actually, if $\kappa=\mu$ we have $\eta(0)=0$ (while in any other case $\eta(z)$ is always positive), and some normal modes have infinitesimal transmissivity. It is then natural that for any nonzero temperature it is not convenient to send energy into these low-capacity modes. More formally, the argument of the positive part in \eqref{nz} in the case $\kappa=\mu$ in $z=0$ is $-N<0$, so for any $N>0$ the positive part must be taken into account.

\section{Conclusions}\label{sec:conc}

In this work we study a model of Gaussian thermal memory channels extending a previous proposal by Lupo {\it et al.} \cite{lupo1,lupo2} in order to incorporate the disturbance of thermal noise. The memory effects imply that successive uses of a channel cannot be
considered independently but they are potentially correlated \cite{memory0,memory1}. In our model this correlation is generated by an internal memory mode which is assumed to be unaccessible by the users of the channel.

Exploiting the factorization into independent normal modes \cite{lupo2} and a recent break-through in the theory of memoryless channels \cite{CC}, we explicitly determine the classical capacity of our memory channel model. We find that, as in the memoryless case, coherent states are sufficient for an optimal coding. However, the associated probability distribution is factorized only in the normal mode decomposition that diagonalizes the channel, so in order to fully exploit its intrinsic memory, the  input signals $\{ a_j\}$ (and consequently their  outputs counterparts) must be correlated. Then the optimal transmission rate of information can still be achieved by independent uses of the channel, but the probability distribution of the physical inputs will not be factorized.

Our results can find applications in  bosonic communication channels with memory effects and affected by a non-negligible amount of thermal noise. In particular low frequency communication devices, {\it e.g.}\  GHz communication systems \cite{microwaves}, THz lasers \cite{tredicucci}, {\it etc.}, are intrinsically subject to black-body thermal noise and thus they fall in the theoretical framework presented in this work.

\section{Acknowledgements}
The authors are grateful to C. Lupo and S. Mancini for useful comments.
G.d.P. thanks A. Tomadin for comments and discussions. This work is partially supported by the EU Collaborative Project TherMiQ (grant agreement 618074).
\clearpage
\begin{minipage}{0.96\columnwidth}
  \centering
  \includegraphics[width=\columnwidth]{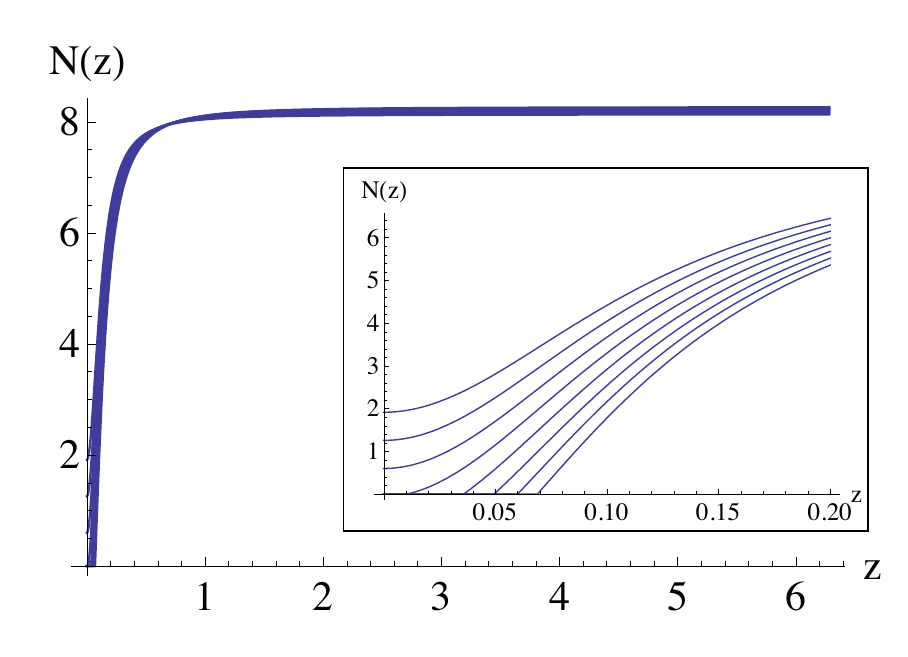}\\
  \figcaption{(Color online): Behavior of the energy density $N(z)$ for $\kappa=0.9$, $\mu=0.8$, $E=8$ and $N$ ranging in steps of 0.1 from top to bottom from 0.5 to 1.2, near to the critical temperature $N_{crit}\sim0.8$. As expected, $N(z)$ is always increasing. If we exclude the region near $z=0$, the functions are almost identical and approach nearly the same constant value for $z\gtrsim1$. Inset:
  Zoom on the region $z\to0$. We can see that above the critical temperature $N(z)$ is zero on a finite interval, while below it $N(z)$ approaches a positive value which strongly depends on the temperature.}\label{n(z)}
\end{minipage}
\\[\intextsep]
\begin{minipage}{\columnwidth}
  \centering
  \includegraphics[width=\columnwidth]{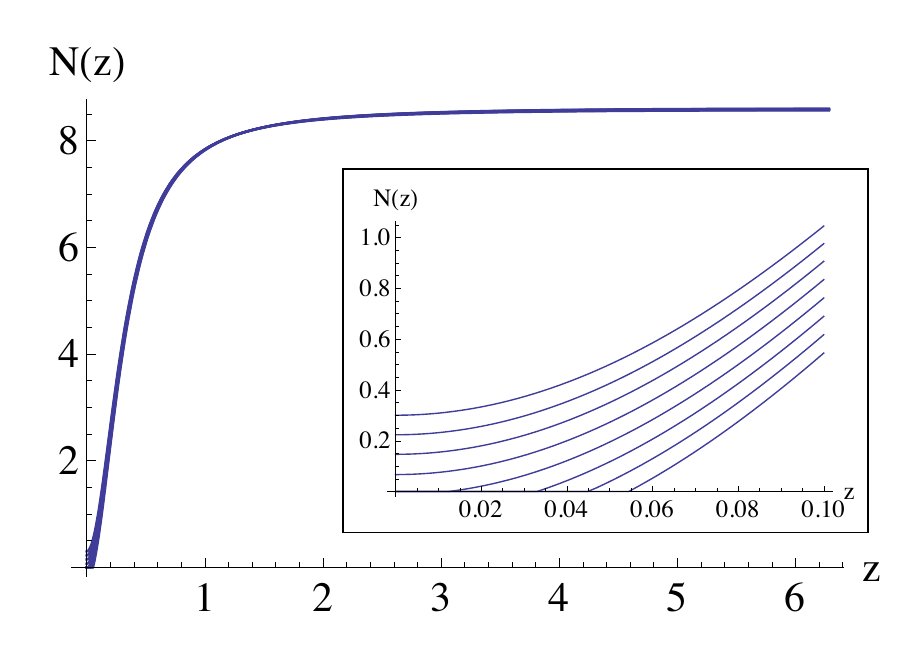}\\
  \figcaption{(Color online): Behavior of the energy density $N(z)$ for $\kappa=1.1$, $\mu=0.8$, $E=8$ and $N$ ranging in steps of 0.1 from top to bottom from 9.4 to 10.1, near to the critical temperature $N_{crit}\sim9.8$. As expected, $N(z)$ is always increasing. If we exclude the region near $z=0$, the functions are almost identical and approach nearly the same constant value for $z\gtrsim1$.
  Inset: Zoom on the region $z\to0$. We can see that above the critical temperature $N(z)$ is zero on a finite interval, while below it $N(z)$ approaches a positive value which strongly depends on the temperature.}\label{n(z)amp}
\end{minipage}
\\[\intextsep]
\begin{minipage}{\columnwidth}
  \centering
  \includegraphics[width=\columnwidth]{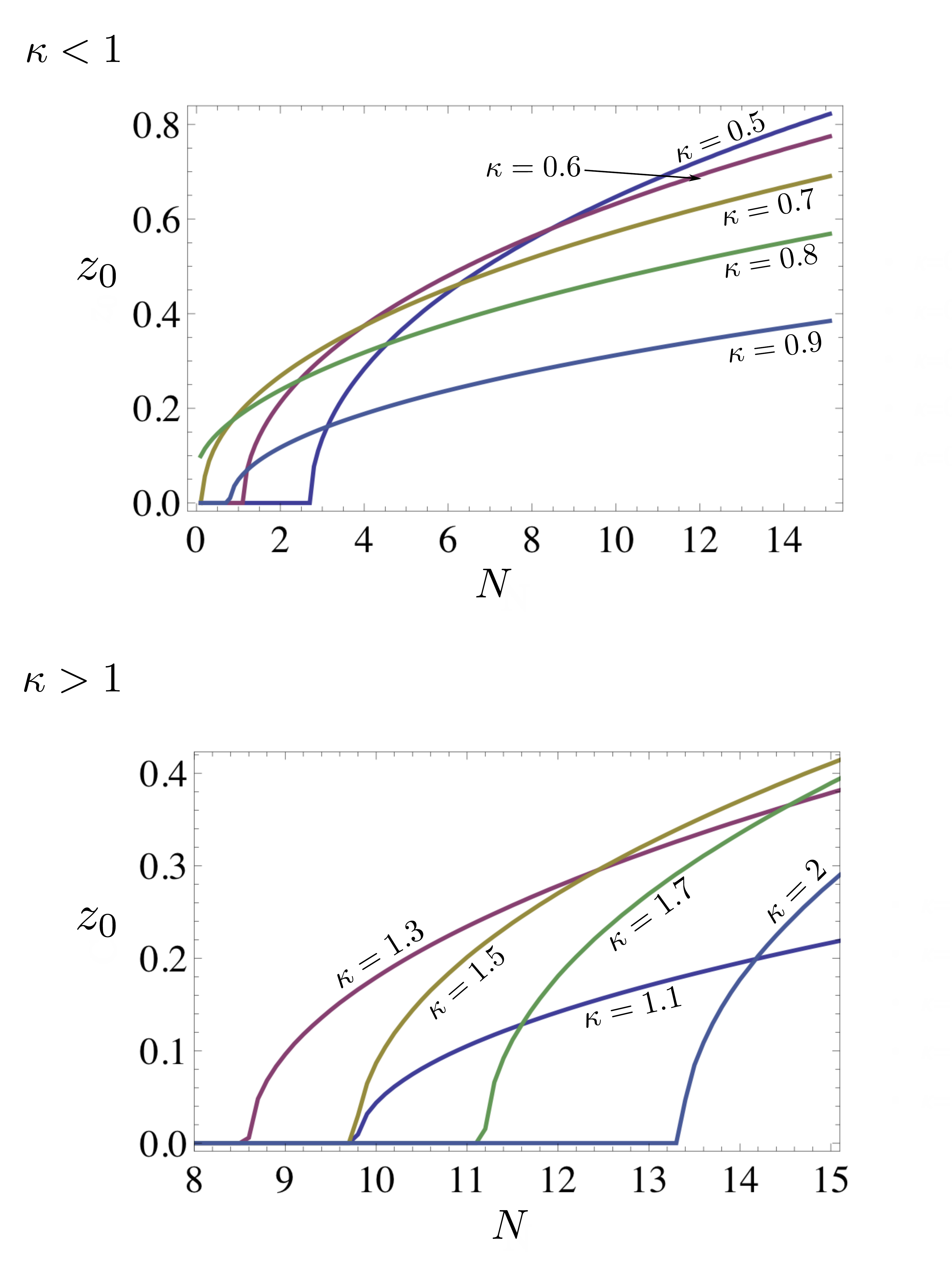}\\
  \figcaption{(Color online): Behavior of the fraction $\frac{z_0}{2\pi}$ ($z_0$ ranges from $0$ to $2\pi$) of unused beam-splitters as a function of the temperature $N$ for $E=8$, $\mu=0.8$ and various values of $\kappa$. At zero temperature ($N=0$) all the beam-splitters are exploited and $z_0=0$; then $z_0$ remains zero up to the critical temperature $N_{crit}$, and grows for $N>N_{crit}$. We notice that for typical values of the parameters $N_{crit}$ is much greater for $\kappa>1$ than for $0<\kappa<1$. In the infinite temperature limit $N\to\infty$ only an infinitesimal fraction of the beam-splitters is used and $z_0$ tends to $2\pi$, even if this is not evident from the plots due to the limited range of $N$.}\label{fig:Z0}
\end{minipage}
\\[\intextsep]
\begin{minipage}{\columnwidth}
  \centering
  \includegraphics[width=\columnwidth]{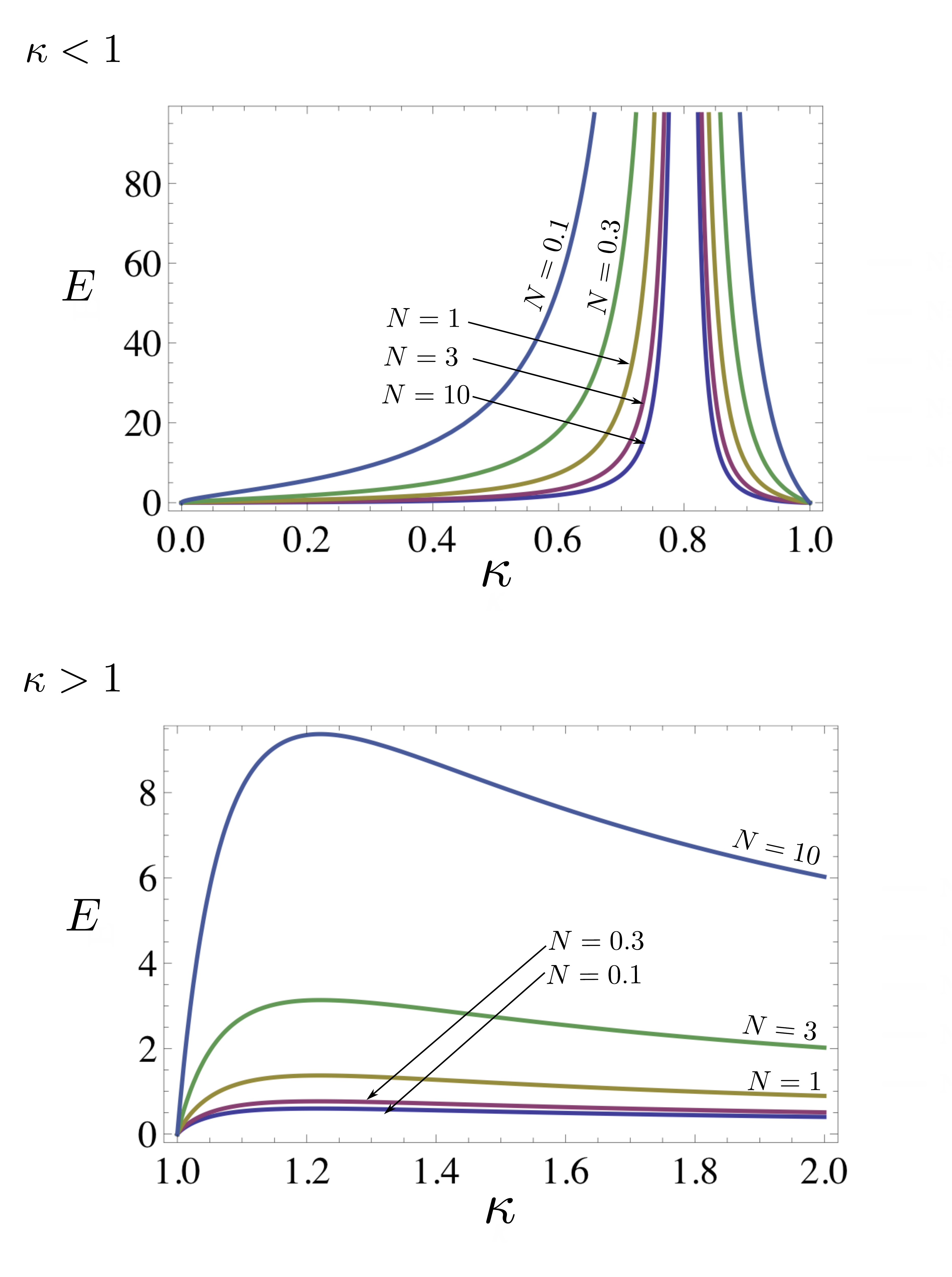}\\
  \figcaption{(Color online): Behavior of the minimal energy for which all the beam-splitters are exploited as a function of $\kappa$ for $\mu=0.8$ and various values of the temperature $N$. As expected, $E$ grows with the temperature, and $E=0$ for $\kappa=0,1$ and $\kappa\to\infty$. In the attenuator case we notice the divergence of $E$ for $\kappa=\mu$ ($=0.8$), due to the fact that $\eta(0)=0$ and for any positive temperature the optimal $N(z)$ must vanish on a finite interval.}\label{fig:Encrit}
\end{minipage}


\clearpage
\begin{thebibliography}{99}


\bibitem{caves}
C. M. Caves and  P. B. Drummond,
%Quantum limits on bosonic communication rates,
Rev. Mod. Phys. {\bf 66},  481 (1994).

\bibitem{holevo} A. S. Holevo, {\it Quantum systems, channels, information. A
mathematical introduction}, (De Gruyter, Berlin-Boston, 2012).

\bibitem{holevowerner} A. S. Holevo and R. F. Werner,
%Evaluating capacities of bosonic Gaussian channels,
Phys. Rev. A {\bf 63},  032312 (2001).

\bibitem{schumacher}
B. Schumacher and M. D. Westmoreland,
%Sending classical information via noisy quantum channels,
Phys. Rev. A {\bf 56}, 131 (1997).



\bibitem{braunstein}
S. L. Braunstein and  P. van Loock,
%Quantum information with continuous variables.
Rev. Mod. Phys. {\bf 77}, 513 (2005).

 \bibitem{gauss1} J. Eisert and M.M. Wolf, Quantum Information with Continous Variables of Atoms and Light, pages 23-42 (2007)

\bibitem{gauss2}  C. Weedbroock, S. Pirandola, R.  Garc{\'i}a-Patr\'on, N. J.  Cerf, T. Ralph, J. H. Shapiro, and S. Lloyd,
%Gaussian Quantum Information.
Rev. Mod. Phys.  {\bf 84},  621 (2012).


\bibitem{conj1}
  V. Giovannetti, A. S. Holevo, and R. Garc{\'i}a-Patr\'on,
 % A solution of the Gaussian optimizer conjecture.
 Eprint  arXiv:1312.2251 [quant-ph].

  \bibitem{conj2}
  A. Mari, V. Giovannetti and A. S. Holevo,
%  Quantum state majorization at the output of bosonic Gaussian channels.
 Eprint  arXiv:1312.3545 [quant-ph].

\bibitem{CC}
V.   Giovannetti,  R. Garc{\'i}a-Patr\'on, N. J. Cerf,  and A. S. Holevo,
%  Ultimate communication capacity of quantum optical channels by solving the Gaussian minimum-entropy conjecture.
Eprint  arXiv:1312.6225 [quant-ph].

\bibitem{strong}
B. R. Bardhan, R. Garc{\'i}a-Patr\'on, M. M. Wilde, and A. Winter,
%Strong converse for the classical capacity of all phase-insensitive bosonic Gaussian channels.
Eprint arXiv:1401.4161 [quant-ph].


\bibitem{memoryrev}
F. Caruso, V. Giovannetti, C. Lupo, and S. Mancini,
Eprint arXiv:1207.5435 [quant-ph].

\bibitem{exp1}
K. Banaszek, A. Dragan, W. Wasilewski, and C. Radzewicz, Phys. Rev.
Lett. {\bf 92}, 257901 (2004); R. Demkowicz-Dobrza\'{n}ski, P.
Kolenderski, K. Banaszek, Phys. Rev. A {\bf 76}, 022302 (2007).


\bibitem{exp2}
E. Paladino, L. Faoro, G. Falci, and R. Fazio, Phys. Rev. Lett. {\bf
88}, 228304 (2002); Y. Hu, Y.-F. Xiao, Z.-W. Zhou, and G.-C. Guo,
Phys. Rev. A  75, 012314 (2007).




\bibitem{memory0}
R. G. Gallager, \emph{Information Theory and Reliable Communication}
(Wiley, New York, 1968).


\bibitem{memory1}
D. Kretschmann and R. F. Werner, Phys. Rev. A {\bf 72}, 062323
(2005).


\bibitem{memory2}
N. Datta and T. C. Dorlas, J. Phys. A: Math. Theor. {\bf 40}, 8147
(2007); A. D' Arrigo, G. Benenti, and G. Falci,  New J. Phys. {\bf
9}, 310 (2007).


\bibitem{memory3}
V. Giovannetti, J. Phys. A {\bf 38}, 10989  (2005).

\bibitem{memory4}
V. Giovannetti and S. Mancini, Phys. Rev. A {\bf 71}, 062304 (2005).

\bibitem{CERF2}
N. J. Cerf, J. Clavareau, C. Macchiavello, and J. Roland, Phys. Rev. A {\bf 72}, 042330 (2005).

\bibitem{lupo1}
C. Lupo, V. Giovannetti and S. Mancini, Phys. Rev. Lett.
\textbf{104}, 030501 (2010).

\bibitem{lupo2}
C. Lupo, V. Giovannetti and S. Mancini, Phys. Rev. A {\bf 82}, 032312 (2010).


\bibitem{OLEG}
O. V. Pilyavets, C. Lupo, and S. Mancini,
	IEEE Trans. Inf. Theory, {\bf 58}, 6126 (2012).
\bibitem{CERFA}
J. Sch\"{a}fer, D. Daems, E. Karpov and N. J. Cerf, Phys. Rev. A {\bf 80}, 062313 (2009).
\bibitem{CERF1}
J. Sch\"{a}fer, E. Karpov and N. J. Cerf, Phys. Rev. A {\bf 84}, 032318 (2011);
J. Sch\"{a}fer, E. Karpov, and N. J. Cerf,
Phys. Rev. A {\bf 85}, 012322 (2012).
\bibitem{additiven1}
C. Lupo, L. Memarzadeh and S. Mancini, Phys. Rev. A {\bf 80}, 042328 (2009).



%%%% applications %%%%

\bibitem{fibers}
N. Gisin, G. Ribordy, W. Tittel, and H. Zbinden, Rev. Mod. Phys.
{\bf 74}, 145 (2002).

\bibitem{waveguides}
S. Tanzilli, {\em  et al.}, Eur. Phys. J. D, {\bf 18}, 155 (2002).

\bibitem{microwaves} C. Lang {\it et al.},  Nature Physics 9, 345-348 (2013)

\bibitem{tredicucci} R. K\"ohler {\it et al.}, Nature 417, 156-159 (9 May 2002)


\bibitem{freespace}
V. W. S. Chan, J. Lightw. Technol., {\bf 24}, 4750 (2006); A.
Fedrizzi, {\em et al.}, Nat. Phys. {\bf 5}, 389 (2009).

%%%%%%%
\bibitem{toeplitz}
R. M. Gray, {\it Toeplitz and Circulant Matrices: A Review}, (Now
Publishers, Norwell, Massachusetts, 2006).


\end{thebibliography}
\end{document}